\documentstyle[11pt]{article}

\begin{document}
\begin{titlepage}
\begin{flushright}
{\small DE-FG05-92ER40717-13 \\
MAD/TH/94-4}
\end{flushright}
\vspace*{12mm}
\begin{center}
               {\LARGE\bf ``Triviality'' and the Perturbative Expansion \\
\vspace*{4mm}
                          in $\lambda\Phi^4$ Theory}

\vspace*{14mm}
{\Large  M. Consoli}
\vspace*{5mm}\\
{\large
Istituto Nazionale di Fisica Nucleare, Sezione di Catania \\
Corso Italia 57, 95129 Catania, Italy}
\vspace*{5mm}\\
and
\vspace*{5mm}\\
{\Large P. M. Stevenson}
\vspace*{5mm}\\
{\large T.~W.~Bonner Laboratory, Physics Department \\
Rice University, Houston, TX 77251, USA}
\vspace{14mm}\\
{\bf Abstract:}
\end{center}

\par The ``triviality'' of $(\lambda\Phi^4)_4$ quantum field theory
means that the renormalized coupling $\lambda_R$ vanishes for
infinite cutoff.  That result inherently conflicts with the usual
perturbative approach, which begins by postulating a {\it non-zero},
cutoff-independent $\lambda_R$.  We show how a ``trivial'' solution
$\lambda_R=0$ can be compatible with the known structure of
perturbation theory to arbitrarily high orders, by a simple
re-arrangement of the expansion.  The ``trivial'' solution reproduces
the result obtained by non-perturbative renormalization of the
effective potential.  The physical mass is finite, while the
renormalized coupling strength vanishes: the two are {\it not}
proportional.  The classically scale-invariant $\lambda \Phi^4$
theory coupled to the Standard Model predicts a 2.2 TeV Higgs, but
does {\it not} imply strong interactions in the scalar sector.

\end{titlepage}

\setcounter{page}{1}

\par  {\bf 1.}
Suppose we accept that the 4-dimensional $\lambda \Phi^4$ theory
is indeed ``trivial'' \cite{froh}, meaning that it has no observable
particle interactions; what is the theory's effective potential?  Since
there are no interactions the effective potential can only be the classical
potential plus the zero-point energy of the free-field fluctuations.
This is the crucial insight of Ref. \cite{cs}:--- {\it for a ``trivial''
theory the one-loop effective potential is effectively exact}.  (A recent
lattice calculation provides striking confirmation of this fact
\cite{agodi}.)

\par  The usual perturbative renormalization \cite{cw} is then not
appropriate because it would spoil this exactness --- it does not properly
absorb the infinities, but merely pushes them into ``higher-order terms''
which are then neglected.  However, it is simple to renormalize the
one-loop effective potential in an exact way \cite{cast,con,iban,cs}.
(This was first discovered in the context of the Gaussian effective potential
\cite{cian,st}.)  The constant background field $\phi$, the argument of
$V_{{\rm eff}}$, requires an infinite re-scaling, but the fluctuation field
$h(x) \equiv \Phi(x) - \phi$ (i.e., the $p_{\mu} \neq 0$ projection of the
field) is not re-scaled \cite{cs}.  The particle mass $m_h$ is related to
the cutoff $\Lambda$ and the bare coupling constant $\lambda=\lambda(\Lambda)$
by
\begin{equation}
\label{mm}
  m_h^2=\Lambda^2 \exp-{{32\pi^2}\over{3\lambda}}.
\end{equation}
Thus, for $m_h$ to remain finite $\lambda$ must vanish like
$1/\ln(\Lambda/m_h)$ in the continuum limit ($\Lambda \to \infty$).
As a consequence one finds that the connected $n$-point functions at
non-zero momentum vanish for $n > 2$, implying no particle interactions;
i.e., ``triviality''.  In particular, the connected 4-point function,
from which one might have hoped to define a renormalized coupling constant
$\lambda_R$, vanishes.

\par  The usual perturbative approach, by contrast, is based on an attempt
to generate a cutoff-independent and {\it non-vanishing} $\lambda_R$.
No meaningful continuum limit is possible in perturbation theory.  In fact,
as discussed by Shirkov \cite{shirkov}, perturbative calculations of the
$\beta$ function up to 5 loops \cite{calc} provide the following results:
In odd orders, $\beta^{{\rm 1-loop}}_{{\rm pert}}$,
$\beta^{{\rm 3-loop}}_{{\rm pert}}$, $\beta^{{\rm 5-loop}}_{{\rm pert}}$
are positive and monotonically increasing.  In even orders
$\beta^{{\rm 2-loop}}_{{\rm pert}}$, $\beta^{{\rm 4-loop}}_{{\rm pert}}$
each have an ultraviolet fixed point, which would imply a finite bare
coupling constant, in contradiction with the rigorous results of
Ref. \cite{froh}.  The magnitude of this spurious fixed point at even
orders appears to decrease to zero with increasing perturbative order.
A Borel re-summation procedure \cite{shirkov,calc} yields a positive,
monotonically increasing $\beta$ function, as in odd orders.  That does
not allow a continuum limit because the renormalized coupling will have
an unphysical Landau pole.

\par  The moral is that only by abandoning, at the start, the vain attempt
to define a non-zero renormalized 4-point function can one obtain a
continuum limit.  In the effective potential analysis \cite{cs} one
actually starts from an approximation scheme (one-loop or Gaussian) in
which, by definition, the shifted field $h(x)$ is non-interacting.  The
resulting effective potential exhibits spontaneous symmetry breaking (SSB)
and allows a continuum limit in which ``dimensional transmutation'' occurs,
with massive particles arising from a scale-invariant bare action.
The renormalization never introduces a ``$\lambda_R$'' but simply
requires the particle mass and $V_{{\rm eff}}$ to be finite.  One finds,
as a consequence, that this renormalization implies ``triviality'' ---
thereby revealing that the original ``approximation'' was effectively
exact.

\par In this Letter we shall follow a different route, considering
the 4-point function of the already {\it massive} theory.   At the
leading-log level, because of the Landau-pole problem, we shall see that
the only possibility for defining a continuum limit of the regularized
theory corresponds to $\lambda_R=0$.  This yields the same relation
(\ref{mm}) as above.  We then show that this solution is compatible with
all orders of sub-leading logarithms.

\vspace*{2mm}
\par  {\bf 2.}
Let us start by defining $\lambda_R$ as the 4-point function
in the limit of zero external momenta (which for massive particles is not
an exceptional point.)  We calculate this in terms of the bare,
cutoff-dependent coupling $\lambda=\lambda(\Lambda)$, taking into account
the basic one-loop bubble of particles with mass $m_h$.  This gives:
\begin{equation}
\label{bub}
\lambda_R = \lambda - b_0 \lambda^2 t,
\end{equation}
where
\begin{equation}
b_0 \equiv \frac{3}{16 \pi^2},
\end{equation}
\begin{equation}
\label{t}
t \equiv \ln (\Lambda/m_h).
\end{equation}
It is evident that the actual expansion is not in powers of $\lambda$ but
rather in powers of $\lambda$ and $t$.  However, one can define a
(perturbative) $\beta$-function that depends on $\lambda$ alone:
\begin{equation}
\label{beta}
\beta_{{\rm pert}} \equiv \Lambda \frac{\partial \lambda}{\partial \Lambda}
= \frac{\partial \lambda}{\partial t}
=b_0\lambda^2+b_1\lambda^3+\ldots .
\end{equation}
[Note that we are defining the $\beta$ function in terms of the cutoff
dependence of the bare coupling constant.  In the conventional
perturbative context this is completely equivalent to the more usual
definition as the renormalization-point dependence of the renormalized
coupling constant.  Since we want to consider the case where $\lambda_R$
vanishes identically the above definition is obviously preferable.]
Formally, by integrating the $\beta$ function one re-sums large
logarithms in the series for $\lambda_R/\lambda$:  The first term takes
into account all leading-log terms, $(b_0 \lambda t)^n$; the second term
accounts for the sub-leading logarithms $\lambda (b_0 \lambda t)^n$, etc..
Using $\beta$ seemingly allows one to relax the requirement
$b_0\lambda t \ll 1$ to just $\lambda \ll 1$.

\par This is powerful magic, and very familiar, but one should be aware
of the hidden assumptions behind it.  The $\beta_{{\rm pert}}$ function
is extracted from an RG equation that is satisfied only in a perturbative
sense, neglecting higher-order terms.  The statement that the leading
term is $b_0 \lambda^2$ is equivalent to assuming that the leading-log
series converges and so can be summed.  That is, one is assuming
$\mid \! b_0\lambda t \! \mid < 1 $.  If the theory were perturbatively
aymptotically free this would create no difficulty, but here $b_0$ is
positive and one has the ``Landau-pole'' problem.  Explicitly, the
solution to $d \lambda/d t = b_0 \lambda^2$, in terms of the boundary
condition at $t=0$, is
\begin{equation}
\label{sol}
\lambda(t) = {{\lambda(0)}\over{1-b_0 \lambda(0) t} }.
\end{equation}
One is forced to identify $\lambda_R=\lambda(0)$ for consistency with
the original equation (\ref{bub}), which is seen as the first two terms
in the infinite expansion of
\begin{equation}
\label{lrp}
\lambda_R=\lambda(0)=
{{\lambda(t)}\over{1+b_0 \lambda(t) t} }.
\end{equation}
One wants to take $\Lambda$, and hence $t$, to infinity, but as
$t$ is increased from zero $\lambda(t)$ grows without bound; indeed
it becomes infinite at $t = 1/(b_0 \lambda_R)$.
Thus, the condition $\mid \! b_0\lambda t \! \mid < 1 $ is inevitably
violated.  No sensible $\Lambda \to \infty$ limit is possible.
This pushes the problem of the continuum limit to the next-to-leading
level.  There, since $b_1<0$, one finds an ultraviolet fixed point; but
this conflicts with the rigorous results of Ref \cite{froh}, and in any
case it disappears at next-to-next-to-leading order.  These results
actually signal the inconsistency, in the $\lambda \Phi^4$ case, of
assuming that the leading-log series can be naively re-summed.

\vspace*{2mm}
\par  {\bf 3.}
Let us re-examine the $\beta$-function approach, relying on just two key
ingredients; (i) a basic equation from which one obtains the $\Lambda$
dependence of $\lambda$, and (ii) the necessity of achieving a continuum
limit $\Lambda \to \infty$.  Our basic equation is Eq. (\ref{bub}) and we
attempt to keep $\lambda_R$ and the physical mass $m_h$ fixed (i.e.
$\Lambda$ independent) while taking the continuum limit $\Lambda\to\infty$.
That is, we demand
\begin{equation}
\label{lr0}
{{d\lambda_R}\over{dt}}= 0,
\end{equation}
which yields
\begin{equation}
\label{full}
{{d\lambda(t)}\over{dt}} - b_0 \lambda^2(t) -
2 b_0 t \lambda(t) {{d\lambda(t)}\over{dt}} = 0.
\end{equation}
In the usual perturbative analysis one would neglect the third term on the
left-hand side of the above equation and arrive at
\begin{equation}
\label{part}
{{d\lambda(t)}\over{dt}} = b_0 \lambda^2(t).
\end{equation}
Seemingly, the neglected term is then ${\cal O}(\lambda(t)^3)$,
justifying the procedure, {\it a posteriori}.  However, one cannot
obtain a continuum limit in this way, as just explained.

\par  If, instead, we {\it do} keep the third term in Eq. (\ref{full})
we obtain
\begin{equation}
\label{fullsol}
{{d\lambda(t)}\over{dt}} = - b_0 \lambda^3(t)
{{1}\over{\lambda(t)-2\lambda_R}}.
\end{equation}
Assuming that $\lambda(t)$ and $\lambda_R$ are both non-negative we find
that Eq. (\ref{fullsol}) has to be studied separately for
$\lambda(t)-2\lambda_R > 0$ and for $\lambda(t)-2\lambda_R < 0$ to preserve
the uniqueness of the solution.  In neither case, however, is a limit
$t\to\infty$ possible if $\lambda_R>0$. The only possibility is associated
with the case $\lambda_R=0$, which gives:
\begin{equation}
\label{betaneg}
{{d\lambda(t)}\over{dt}}=-b_0\lambda^2(t),
\end{equation}
\begin{equation}
\label{lamus}
   \lambda(t)={{1}\over{b_0 t}}.
\end{equation}
Thus, now we find a {\it negative} $\beta$ function, giving a bare coupling
constant that tends to zero in the continuum limit.  Eq. (\ref{lamus})
is precisely the relation (\ref{mm}), obtained from the effective-potential
analysis of the massless theory \cite{cs}.  [The above explicitly answers
the objection of Ref. \cite{at}: our $\beta$ function is not, of course,
$\beta_{{\rm pert}} + {\mbox{{\rm (non-perturbative corrections)}}}$; it is
simply the {\it right} $\beta$ function for achieving a continuum limit.]

\vspace*{2mm}
\par  {\bf 4.}
To discuss higher orders it is convenient to introduce the variable
\begin{equation}
\label{x}
     x=b_0 \lambda(t) t.
\end{equation}
The basic one-loop correction, Eq. (\ref{bub}), then has the form
$\lambda^{(0)}_R=\lambda(t) (1-x)$.  Explicit calculation of the
higher-order leading-logarithmic corrections to this formula would
of course give $\lambda_R = \lambda(t)(1-x+x^2-x^3+\ldots)$, in
agreement with a formal expansion of $\lambda_R = \lambda(t)/(1+x)$
(Eq. (\ref{lrp})).  However, that expression represents a re-summation
of the geometric series that is only valid if $\mid \! x \! \mid < 1$.
Our solution, Eq. (\ref{lamus}), is $x=1$ with $\lambda_R=0$.
It is easy to see that this can be a solution to arbitrarily high order
if we rearrange the perturbative expansion suitably.  We can
view the higher-order diagrams as modifying, and multiplicatively
renormalizing, $\lambda_R^{(0)}$ rather than $\lambda(t)$.  In a sense,
this makes the effective expansion parameter $x^n(1-x)$ rather than
$x^n$ itself.  For $x \ll 1$ this would make essentially no difference,
of course.  It produces a sequence of approximations (for $N=0,1,2,\ldots$)
of the form
\begin{equation}
\label{lamn}
\lambda^{(N)}_R=\lambda(t)(1-x)(1+x^2+x^4+...x^{2N})=\lambda(t)
{{1-(x^2)^{N+1}}\over{1+x}}
\end{equation}
which, for any $N$, gives
\begin{equation}
\label{zero}
\left. \lambda^{(N)}_R \right|_{x=1} = 0.
\end{equation}
Note that the limits $x\to 1$ and $N\to\infty$ do not commute.
Indeed, for any $x<1$ one has
\begin{equation}
\label{xl1}
\lambda_R= \lim_{N\to\infty}~\lambda^{(N)}_R(x)={{\lambda(t)}\over{1+x}},
\end{equation}
whose $x \to 1$ limit is
$\lambda_R=\frac{1}{2} \lambda(t)$, whereas we have
\begin{equation}
\label{zeroag}
\lambda_R=\lim_{N\to\infty}~\lim_{x\to 1}~\lambda^{(N)}_R(x)=0,
\end{equation}
yielding again Eqs. (\ref{betaneg}, \ref{lamus}).

\par This procedure can be extended to include all orders of sub-leading
logarithms.  The essential point is that any sub-leading-log term $A$
appearing at some order in $\lambda$ will itself be modified in subsequent
orders by a series of leading-log corrections, $A(1-x+x^2-\ldots)$, and
so is multiplied by a $\lambda^{(N)}_R$ factor.  For instance, the sequence
of approximations
\begin{equation}
\label{seq}
\lambda^{(N,M+1)}_R={ { \lambda^{(N)}_R }\over{
1-c\lambda^{(N)}_R
\ln{{\lambda^{(N,M)}_R(1+c\lambda(t))}\over{\lambda(t)(1+c\lambda^{(N,M)
}_R)}}}},
\end{equation}
with $c \equiv b_1/b_0$, contains, in the limit $N\to\infty$, $M\to\infty$,
all the leading and next-to-leading corrections to the zero-momentum
coupling, $\lambda_R$.  One can see this as follows.  For
$\mid \! x \! \mid < 1$ the above sequence corresponds to an iterative
solution of the implicit equation
\begin{equation}
\label{ff}
\lambda_R=\frac{ \lambda_{\l\l} }{ 1-c\lambda_{\l\l}
\ln{ \frac{\lambda_R(1+c\lambda(t))}{\lambda(t)(1+c\lambda_R)} } }
\end{equation}
where $\lambda_{\l\l} = \lim_{N \to \infty} \lambda^{(N)}_R$ is the
leading-log solution, which is $\lambda_{\l\l} = \lambda(t)/(1+x)$ for
$\mid \! x \! \mid < 1$.  It is then straightforward to
check that for $\lambda_R$ to be cutoff independent one requires
$\lambda(t)$ to satisfy
\begin{equation}
\label{2loop}
{{d\lambda(t)}\over{d t}}=
b_0 \lambda(t)^2(1 + c \lambda(t)),
\end{equation}
which is the two-loop perturbative $\beta$ function.
However, for $x=1$ the sequence (\ref{seq}) gives identically
\begin{equation}
\label{zerost}
\lambda_R=\lim_{N\to\infty}~\lim_{M\to\infty}~\lim_{x\to 1}~
\lambda^{(N,M)}_R(x)=0.
\end{equation}
and the associated relations (\ref{betaneg}, \ref{lamus}).
[Note that the re-summations producing the logarithmic term in the
denominator of Eq. (\ref{seq}) can be performed consistently even when
$x \to 1$ as $t \to \infty$ since both $\lambda^{(N)}_R$ and
$\lambda(t)$ vanish in that limit.]

\par  In other words, we have exploited the fact that the structure of the
sub-leading logarithms can be inferred from the usual perturbative $\beta$
function, which just represents a formal re-summation of those terms.
However, that re-summation is valid only for $\mid \! x \! \mid < 1$.
The sub-leading logarithmic structure itself, though, when examined
{\it at} $x=1$, is consistent to all orders with the `trivial' solution
$\lambda_R=0$.  The point is that all sub-leading corrections are
themselves multiplied by a $\lambda^{(N)}_R$ factor.

\par  Of course all of the above is open to the objection that we are
merely re-arranging the terms of a divergent series.  There is no
defence to this charge.  Our point, though, is that the conventional
procedure, re-summing leading logs to all orders, then sub-leading logs,
etc., is itself a re-arrangement of a divergent series.  Moreover,
because of the Landau pole, one is forced into a region with $x \ge 1$
where this re-arrangement is highly dubious because the sub-series
being re-summed are themselves divergent.

\vspace*{2mm}
\par  {\bf 5.}
To further illustrate our point we give a concrete example.
This is not meant to represent how things actually work in
$\lambda \Phi^4$ theory, but merely to reinforce the point that the
conventional procedure, although sanctified by time and custom, can
in fact give the wrong answer.  Consider the mathematical example in
which $\lambda_R$ and the bare $\lambda$ are related by:
\begin{equation}
\label{orig}
\lambda_R = \lambda (1-x) \sum_{n=0}^{\infty} g_n(\delta) x^{2n},
\end{equation}
where $\delta$ is a parameter that vanishes in the infinite-cutoff limit
(say, as $1/\Lambda$).  If the coefficients
$g_n(\delta)$ all become unity in the infinite-cutoff limit (i.e., as
$\delta \to 0$), then this reproduces the leading-log series
$\lambda_R =\lambda(1-x+x^2-\ldots)$.
However, suppose that in the double limit $\delta \to 0$ and
$n \to \infty$
\begin{equation}
g_n(\delta) \to \left\{
\begin{array}{ll}
1 & \quad\quad {\rm if} \,\, n\delta \ll 1, \\
0 & \quad\quad {\rm if} \,\, n\delta \gg 1.
\end{array}
\right.
\end{equation}
This could happen in many ways; e.g. $g_n(\delta) = (1-\delta)^n$ or
$g_n(\delta) = 1/(1+n! \delta^n)$.
While all $g_n$'s become unity as $\delta \to 0$ for any finite $n$,
we must be careful because our series involves infinitely large $n$.
For any finite $\delta$, no matter how small, the $g_n$ coefficients at
very large $n$ ($n > 1/\delta$) become much less than unity.  Thus, for
$\delta \to 0$ we have
\begin{eqnarray}
\lambda_R  & \sim & \lambda (1-x) \left( \sum_{n=0}^{1/\delta-1} x^{2n}
+ \sum_{n=1/\delta}^{\infty} g_n(\delta) x^{2n} \right)
\nonumber \\
& \sim & \lambda (1-x) \left( \frac{(1-x^{2/\delta})}{(1-x^2)} + R(x) \right)
\nonumber \\
\label{gen}
& \sim & \lambda \frac{(1-x^{2/\delta})}{(1+x)} + \lambda (1-x) R(x),
\end{eqnarray}
The remainder term $R(x)$ is a series beginning at order $x^{2/\delta}$
with a radius of convergence greater than unity, and so is non-singular
at $x=1$.

\par     For $\mid \! x \! \mid < 1$ one has
$x^{2/\delta} \to x^{\infty} \to 0$, and $R(x) \to 0$, so that:
\begin{equation}
\lambda_R  \sim \frac{\lambda}{(1+x)},
\end{equation}
which is the usual perturbative relationship, at the leading-log level.
In this case the subtlety about the $\delta \to 0$ limit of the $g_n$'s
is irrelevant.

\par   However, for $x=1$ the last equation is {\it not} valid.  One then
has $x^{2/\delta} = 1^{\infty} =1$ in (\ref{gen}), and so $\lambda_R =0$.
This is obvious from the $(1-x)$ factor in the original equation,
(\ref{orig}).  The point is that the $g_n x^{2n}$ series does {\it not}
generate a $1/(1-x)$ factor to cancel it.  Thus, this example admits
the ``triviality'' solution, $\lambda_R  = 0$ and $x = 1$, associated with
the {\it negative} $\beta$ function of Eq. (\ref{betaneg}).

\vspace*{2mm}
\par  {\bf 6.}
In conclusion, we have presented a simple rearrangement
procedure which reproduces the full perturbative expansion at
arbitrarily high orders and is valid in the full range $x \le 1$
($x \equiv b_0 \lambda t$).  It is based on the simple remark that
$x^n(1-x)$ is a more suitable expansion parameter than $x^n$ itself.
The assumption $x<1$ allows one to re-sum the various sub-series and
leads to the conventional results.  However, one cannot then obtain
any consistent continuum limit, and moreover one cannot avoid being
dragged into a region with $x \ge 1$, invalidating the original
assumption.  However, if the continuum limit is governed by $x \to 1$,
then the condition $\lambda_R=0$ holds to all orders in this modified
expansion.  This solution is entirely consistent with the
``triviality'' found in mathematically rigorous analyses \cite{froh}.
It is also entirely consistent with the effective-potential analysis
\cite{cs}, which is based on the very physical consideration that the
effective potential of a ``trivial'' theory is just the classical
potential plus the zero-point energy of the free-field  fluctuations.

\par  Our results here {\it prove} nothing, since we start from an
inherently divergent Feynman-diagram expansion.  However, they
do provide a way to understand how ``triviality'' can be consistent
with a seemingly highly non-trivial perturbative structure.

\par The consequences of our picture are substantial, and are discussed
in more detail in Ref. \cite{cs}.  Although the $\lambda \Phi^4$ theory
is ``trivial'' (i.e., has non-interacting particles), it has SSB.
When coupled to the Standard Model --- and the gauge and Yukawa
interactions may be treated as small perturbations --- it leads to the
Higgs-Kibble mechanism in the usual way.  In the theoretically most
attractive classically-scale-invariant case, one finds \cite{con,cs} the
relation $m_h^2 = 8 \pi^2 v^2$, where $v$ is the renormalized expectation
value of the scalar field, known from the Fermi constant to be 246 GeV.
Thus, one predicts a 2.2 TeV Higgs boson \cite{con,cs}.  In the
perturbative picture $m_h$ would be proportional to $\lambda_R$, but
in the ``trivial'' solution $m_h$ and $\lambda_R$ are quite distinct
quantities: the former remains finite while the latter vanishes \cite{huang}.
Thus, in spite of the large Higgs mass, the Higgs/longitudinal-$W,Z$ sector
in our picture is {\it not} strongly interacting.  Indeed, the interactions
in this sector are of electroweak strength, and would vanish if the gauge
and Yukawa couplings were turned off.

\vspace*{2mm}
\begin{center}
{\bf Acknowledgements}
\end{center}
\par M.~C. would like to thank D. Shirkov for very useful discussions.
P.~M.~S. thanks the Physics Department of the University of
Wisconsin--Madison for their hospitality.

This work was supported in part by the U.S. Department of Energy under
Grant No. DE-FG-92ER40717 (Rice) and Contract No. DE-AC02-76ER-00881
(Wisconsin).

\end{document}